\documentstyle[11pt,epsfig,rotate]{article}
\begin{document}
\def\Tr{\,{\rm Tr}\,}
\def\beq{\begin{equation}}
\def\eeq{\end{equation}}
\def\beqa{\begin{eqnarray}}
\def\eeqa{\end{eqnarray}}
\begin{titlepage}
\vspace*{-1cm}
\noindent
\phantom{bla}
\\
\vskip 2.0cm
\begin{center}
{\Large {\bf Electromagnetic Corrections to $K \to \pi\pi$ II -- 
Dispersive Matching}}
\end{center}
\vskip 1.5cm
\begin{center}
{\large Vincenzo Cirigliano$^a$, John F. Donoghue$^b$ 
and Eugene Golowich$^b$} \\
\vskip .15cm
$^a$ Dipartimento di Fisica dell'Universit\`a and I.N.F.N. \\
Via Buonarroti,2 56100 Pisa (Italy) \\
vincenzo@het2.physics.umass.edu \\

\vskip .15cm

$^b$ Department of Physics and Astronomy \\
University of Massachusetts \\
Amherst MA 01003 USA\\
donoghue@physics.umass.edu \\
gene@physics,mass.edu \\

\vskip .3cm
\end{center}
\vskip 1.5cm
\begin{abstract}
\noindent
We express the leading electromagnetic corrections 
in $K \to \pi \pi$ as integrals over the virtual 
photon squared-momentum $Q^2$.  The high $Q^2$ behavior 
is obtained via the operator product expansion.  The 
low $Q^2$ behavior is calculated using chiral perturbation 
theory. We model the intermediate $Q^2$ region using 
resonance contributions in order to enforce the matching of 
these two regimes. Our results confirm our previous estimates that
the electromagnetic corrections provide a reasonably small shift in
the $\Delta I =3/2$ amplitude. 
\end{abstract}
\vfill
\end{titlepage}

\section{\bf Introduction}

\begin{figure}
\vskip .1cm
\hskip 2.8cm
\epsfig{figure=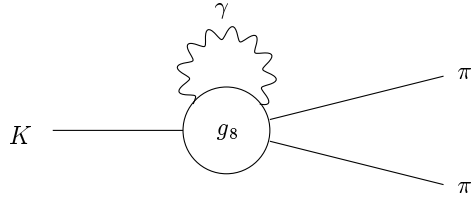,height=1.0in}
\caption{Leading electromagnetic correction to $K \to \pi \pi$.\hfill 
\label{fig:f1}}
\end{figure}

In a previous publication~\cite{cdg2}, we calculated 
the leading electromagnetic corrections to $K \to \pi\pi$ nonleptonic 
decays within chiral perturbation theory 
(ChPT).\footnote{By `leading' is meant the component which arises from 
electromagnetic corrections to the (large) $\Delta I = 1/2$ 
amplitude ({\it cf} Fig.~\ref{fig:f1}).  
This approach will be followed here.}  The only 
hadronic degrees of freedom were the pseudoscalar mesons.  
Loop integrals were analyzed in terms of dimensional 
regularization, and counterterm amplitudes were introduced 
to cancel all divergences.  The finite counterterms parameterize 
the short distance effects of heavy degrees of freedom.  Our 
ChPT analysis yielded effects which were estimated to be
at the several per cent level.
Unfortunately, due to the presence of many unknown finite 
counterterms the results were accompanied by error bars as 
large as the signals.  
 
\subsection{The Method of Dispersive Matching}
In this paper we extend the previous calculation to higher 
energies by using 
a `dispersive matching' approach.  Here, active degrees of 
freedom include not only the ground state pseudoscalar 
mesons but also the spin-zero and spin-one low-lying meson 
resonances.  Amplitudes are expressed as integrals over 
the virtual photon euclidean squared-momentum $Q^2$. 
Within chiral perturbation theory, the $Q^2$ integral is 
regulated dimensionally, and unknown constants are introduced
to parameterize the contributions from intermediate and high energy.
In contrast, the dispersive matching approach is an attempt to 
construct an intermediate energy contribution that 
sucessfully interpolates between the low and high energy regions.
This allows the full $Q^2$ integral to be calculated. 

The $K \to \pi \pi$ amplitude with EM 
interactions present is given generally 
to order $e^2$ by 
\begin{equation}
{\cal A}_i = e^2 \ \int \, d^{4} q \ D_{\mu \nu} (q) \ W_i^{\mu \nu}
(q,p) \qquad (i = +-,00,+0) \ \ , 
\end{equation}
where $D^{\mu \nu} (q)$ is the photon propagator and 
$W_i^{\mu \nu}(q,p)$ describes the scattering $\gamma K(p) \to 
\gamma \pi \pi$.   Rotation to euclidean momentum 
space followed by evaluation of the angular integral yields the 
equivalent form 
\begin{equation}
{\cal A}_i = \frac{\alpha}{4 \pi} \ \int \, d Q^2 \ W_i(Q^2) 
\qquad (i = +-,00,+0) \ \ . 
\label{a2}
\end{equation}
To determine the $K \to \pi \pi$ amplitudes to order $e^2 p^2$
in chiral power counting requires knowing $W_{\mu
\nu} \, (q,p)$ (or equivalently $ W \, (Q^2) $) at
order $p^2$ and for all values of $Q^2$.  We have rigorous 
information on $W_{\mu \nu} \, (q,p)$ only in the two asymptotic 
regimes of (i) low $Q^2$, where ChPT provides the appropriate 
couplings, and (ii) high $Q^2$, where the quark
degrees of freedom couple to the photon according to the standard
electroweak theory. Our goal will be to match these regions.

\begin{figure}
\vskip .1cm
\hskip 1.8cm
\epsfig{figure=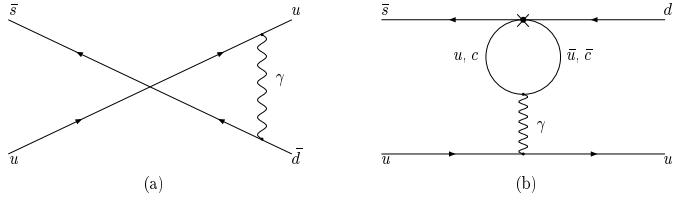,height=1.0in}
\caption{High-$Q^2$ electroweak dynamics of quarks.\hfill 
\label{fig:f2}}
\end{figure}

Consider the process of building $W \, (Q^2)$ 
from the low-$Q^2$ end.  In principle we can use ChPT to generate 
$W_{\mu \nu} \, (q,p)$ to order $p^2 Q^{2n}$.  At the lowest 
energies, the dominant contribution is from the ground state mesons. 
At intermediate energies resonance degrees of freedom 
become active, and we can use effective lagrangians to 
describe their interactions.  Such resonance contributions serve 
to soften the polynomial behavior as $Q^2$ increases.~\cite{cdg}  
Eventually, the low/intermediate $Q^2$ description is matched to the 
high-$Q^2$ effects of Fig.~\ref{fig:f2}.  

Neither of the short 
distance contributions depicted in Fig.~2 plays a dynamical role 
in the radiative problem but for different reasons.  
The process of Fig.~\ref{fig:f2}(a) leads merely to an overall 
shift in the strength of the weak interaction but does not give 
rise to mixing between the isospin amplitudes.  The electroweak 
penguin operators of Fig.~\ref{fig:f2}(b) do 
contribute to $K \to \pi \pi$ decay, but are found to be quite 
small~\cite{cdg} and so are neglected in the work reported here.
Physics of the low-to-intermediate $Q^2$ region is therefore 
the dominating influence in our calculation.  

In Section~2, we define the various interaction lagrangians 
which are needed in the course of the calculation.  We present 
a detailed account of the calculational program in Section~3, 
from its content through to the results and some phenomenological 
implications. We pay particular attention to the uncertainties inherent in
our calculation, and attempt to provide realistic error estimates.
Final remarks appear in Section~4.  

\section{\bf Effective Lagrangians}
Our starting point will be a tree level calculation of the 
$\{ W_i^{\mu\nu} (Q^2)\}$ including as intermediate states the ground state mesons 
and the low lying resonances.  Their 
interactions are dictated by the lowest order chiral lagrangians (of order
$p^2$). Specifically, in the resonance sector we include the 
vector (V), axialvector (A), scalar (S) and pseudoscalar (P) octets 
and the scalar (S1) and pseudoscalar (P1) singlets.

\subsection{Ground State Mesons}
The $|\Delta S| = 1$ octet lagrangian which governs the 
spinless ground state mesons begins at chiral order $p^2$, 
\beq
{\cal L}_8^{(2)} \ = \ g_8 \Tr \left( \lambda_6 D_\mu U D^\mu U^\dagger
\right) \ \ ,
\label{r0}
\eeq
with $g_8 \simeq 6.7 \cdot 10^{-8}~F_\pi^2$ and 
$U \equiv \exp{(i \lambda \cdot \Phi)}$.
The corresponding $\Delta S = 0$ strong/electromagnetic 
lagrangian is 
\beq
{\cal L}^{(2)}_{\rm str} = {F_\pi^2 \over 4} 
\Tr \left( D_\mu U D^\mu U^\dagger
\right) + {F_\pi^2 \over 4} \Tr \left( \chi U^\dagger 
+ U \chi^\dagger \right) \ \ ,
\label{r0a}
\eeq
where $\chi \equiv 2 B_0 ~diag (m_u , m_d , m_s)$.
and $D_\mu U \equiv \partial_\mu U + i e [Q,U] A_\mu$, with 
$A_\mu$ being the photon field.

\subsection{Spin-0ne Resonances}
The spin-one vector and axialvector resonances which 
enter our calculation are represented respectively by the field 
matrices $R_{\mu\nu} = V_{\mu\nu} , A_{\mu\nu}$, 
\beq
V_{\mu\nu} = \left[ \begin{array}{ccc}
\rho^0 /\sqrt{2} + \omega_8 / \sqrt{6} & \rho^+ & K^{*+} \\
\rho^- & -\rho^0 /\sqrt{2} + \omega_8/ \sqrt{6} & K^{*0} \\
K^{*-} & {\overline K}^{*0} & - 2 \omega_8 /\sqrt{6} 
\end{array} \right]_{\mu\nu} 
\label{r0b}
\eeq
and 
\beq
A_{\mu\nu} = \left[ \begin{array}{ccc}
a_1^0 /\sqrt{2} + f_1 / \sqrt{6} & a_1^+ & K_1^{+} \\
a_1^- & - a_1^0 /\sqrt{2} + f_1 / \sqrt{6} & K_1^{0} \\
K_1^{-} & {\overline K}_1^{0} & - 2 f_1 /\sqrt{6} 
\end{array} \right]_{\mu\nu} \ \ .
\label{r0c}
\eeq
The normalization of $R_{\mu\nu}$ is given by 
\beq
\langle 0 | R_{\mu\nu} | R (p, \lambda ) \rangle = 
 {i \over M_R} \left( p_\mu \epsilon_\nu (p, \lambda) - 
p_\nu \epsilon_\mu (p, \lambda) \right) \ \ .
\label{r0d}
\eeq

Analogous to interactions among the spinless ground-state mesons, 
interactions of the resonances are likewise given 
in terms of effective lagrangians.~\cite{weakres}   
For $\Delta S  = 0$ vertices we have 
\beq
{\cal L}_{\rm str}^{(R)} = { F_{\rm V} \over 2 \sqrt{2}} 
\Tr \left( V_{\mu\nu} f_+^{\mu\nu} \right) + 
i { G_{\rm V} \over 2 \sqrt{2}} \Tr \left( 
V_{\mu\nu} u^\mu u^\nu \right) 
+  { F_{\rm A} \over 2 \sqrt{2}} \Tr \left( 
A_{\mu\nu} f_-^{\mu\nu} \right) \ \ , 
\label{r3}
\eeq 
where 
\beqa
\begin{array}{l}
U = u u \ , \\
f_\pm^{\mu\nu} = u^\dagger F^{\mu\nu} u \pm 
u F^{\mu\nu} u^\dagger  \ , 
\end{array}
\qquad 
\begin{array}{l}
u_\mu = i u^\dagger D_\mu U u^\dagger \ , \\
F^{\mu\nu} = eQ (\partial^\mu A^\nu - \partial^\nu A^\mu) 
\ \ .
\end{array}
\label{r4}
\eeqa
The couplings $F_V , G_V , F_A$ have the numerical values~\cite{dp}
\beq
F_V \simeq 0.154~{\rm GeV}\ , \qquad 
G_V \simeq {F_\pi^2 \over F_V}\ , \qquad 
F_A \simeq \left( F_V^2 - F_\pi^2 \right)^{1/2} \ \ .
\label{r5}
\eeq
Although the effective lagrangian used to describe $|\Delta S| = 1$ 
interactions of the resonances is given most generally by~\cite{ekw} 
\beq
{\cal L}_{\rm R} \ = \ \sum_{k = 1}^{10} \ g^{\rm (R)}_k ~K^{\rm (R)}_k
\ \ , 
\label{r1}
\eeq
only four of the ten possible operators 
are relevant to our $K \to \pi \pi$ analysis, 
\beqa
\begin{array}{l}
K^{\rm (R)}_1 = \Tr \left( \Delta [R_{\mu\nu} , f^{\mu\nu}_+ ]_+ 
\right) \ ,  \\
K^{\rm (R)}_5 = i \Tr \left( \Delta [R_{\mu\nu} , 
[u^\mu , u^\nu ] ]_+ \right) \ , 
\end{array}
\qquad 
\begin{array}{l}
K^{\rm (R)}_2 = \Tr \left( \Delta [R_{\mu\nu} , f^{\mu\nu}_- ]_+ 
\right) \ , \\
K^{\rm (R)}_6 = i \Tr \left( \Delta u_\mu R_{\mu\nu} 
u_\nu \right) \ \ , 
\end{array}
\label{r2} 
\eeqa
where $\Delta = u^\dagger \lambda_6 u$.  Use of the 
$ \{ K^{\rm (R)}_k \}$ introduces eight couplings 
$\{ g^{\rm (R)}_k \}$ ($R = V,A$ and $k = 1,2,5,6$) into the 
calculation.  It is convenient to convert these to 
dimensionless quantities, 
\beq
g_k^{\rm (V)} = {g_8 F_V \over F^2} {\bar g}_k^{\rm (V)} \ ,
\qquad 
g_k^{\rm (A)} = {g_8 F_A \over F^2} {\bar g}_k^{\rm (A)} \ ,
\quad (k = 1,2,5,6) \ \ .
\label{r6}
\eeq

\subsection{Spin-zero Resonances}
Finally, we list effective lagrangians for the 
spinless resonances, including the octet scalars $S$, 
the singlet scalar $S_1$ and their pseudoscalar 
analogs $P$ and $P_1$.   We begin with the 
strong lagrangians, 
\beqa
{\cal L}_{\rm str}^{\rm (scalar)} &=&  c_{d} 
\Tr \left( S u_{\mu} u^{\mu} \right) +
c_{m} \Tr \left( S \chi_{+} \right) \nonumber \\
& & \phantom{xxxxx} +  \tilde{c}_{d} S_{1} \Tr 
\left( u_{\mu} u^{\mu} \right) +
 \tilde{c}_{m} S_{1} \Tr \left( \chi_{+} \right) \ \ , 
\nonumber \\
{\cal L}_{\rm str}^{\rm (pseudo)} &=&  i d_{m} 
\Tr \left( P \chi_{-} \right) + 
 i \tilde{d}_{m} P_{1} \Tr \left( \chi_{-} \right) \ \ ,
\eeqa
where $\chi_\pm \equiv u \chi^\dagger u \pm u^\dagger \chi u^\dagger$.
The weak lagrangian for the octet spinless resonances is
\begin{equation} 
 {\cal L}_{\rm wk}^{\rm (octet)} \ = \ \sum_{i = 1}^{6} \, 
g^{i}_{S} K_{i}^{S} \ 
+ \ \sum_{i = 1}^{4} \, g^{i}_{P} K_{i}^{P} \ \ ,
\end{equation}
where 
\beq
\begin{array}{l}
K_{1}^{S} =  \Tr \left( \Delta \, [ S, \chi_{+} ]_+ \right) \ ,\\
K_{3}^{S} = \Tr \left( \Delta \, [ S, \chi_{-} ]_+ \right) \ ,\\
K_{5}^{S} = \Tr \left( \Delta \, u^{\mu} \right) \cdot 
\Tr \left( u_{\mu} S \right) \ ,
\end{array}
\qquad 
\begin{array}{l}
K_{2}^{S} = \Tr \left( S \Delta \, \right) \cdot 
\Tr \left( \chi_{+} \right) \ ,\\
K_{4}^{S} = \Tr \left( \Delta \, [ S, u_{\mu} u^{\mu} ]_+ \right) \ ,
\\
K_{6}^{S} = \Tr \left( \Delta \, S \right) \cdot 
\Tr \left( u_{\mu} u^{\mu}  \right) \ ,
\end{array}
\eeq
and 
\beq
\begin{array}{l}
K_{1}^{P} = i  \Tr \left( \Delta \, [P, \chi_{-} ]_+ \right) \ ,\\
K_{3}^{P} = i \Tr \left( \Delta \, \left[ \chi_{+}, P \right] \right)
\ ,
\end{array}
\qquad 
\begin{array}{l}
K_{2}^{P} = i \Tr \left( \Delta \, P \right) \cdot \Tr 
\left( \chi_{-} \right) \ ,  \\
K_{4}^{P} = i \Tr \left( \Delta \, \left[ P, u_{\mu} u^{\mu} \right] 
\right) \ \ .
\end{array}
\eeq
The weak lagrangian for the singlets is given by 
\begin{equation}
{\cal L}_{\rm wk}^{\rm (singlet)} \ = \ \tilde{g}^{1}_{P}
\tilde{K}_{1}^{P} \ + \ 
\sum_{i = 1}^{2} \ \tilde{g}^{i}_{S} ~
\tilde{K}_{i}^{S} \ \ ,
\end{equation} 
with 
\beq
\tilde{K}_{1}^{S} =  S_{1} \Tr \left( \Delta \, \chi_{+} \right) ~ ,  \quad 
\tilde{K}_{2}^{S} =  S_{1} \Tr \left( \Delta \, u_{\mu} u^{\mu} 
\right) ~ , \quad 
\tilde{K}_{1}^{P} = i P_{1} \Tr \left( \Delta \, \chi_{-} \right) \ .
\eeq

\section{Details of the calculation}\label{details}

\begin{figure}
\vskip .1cm
\hskip 1.8cm
\epsfig{figure=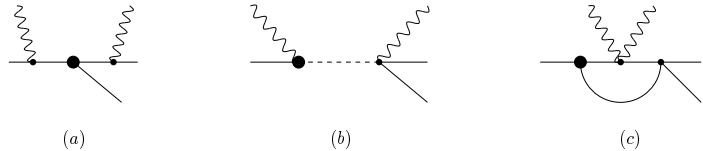,height=0.75in}
\caption{Contributions to $W_i^{\mu\nu}$: (a) Born, 
(b) resonance, (c) loop.\hfill 
\label{fig:f3}}
\end{figure}

We have at hand the tools to construct reliable expressions 
at low and intermediate $Q^2$ for the $\{ W_{i} (Q^2)\}$ 
functions of Eq.~(\ref{a2}).  The two major components 
will be: 

\begin{enumerate} 
\item Tree diagrams (Figs.~\ref{fig:f3}(a),(b)) 
involving exchange of the ground state pseudoscalar
mesons (Born terms) and of resonances: 
\subitem Within chiral perturbation theory, the vertices in the 
tree diagrams are described by point-like couplings at 
leading order. However, in QCD we know that the couplings
fall off at higher $Q^2$. In order to incorporate this feature, 
we model form factor corrections to the Born terms 
with vector resonance contributions.  The set of Born diagrams 
(together with insertions of meson form factors) is free of unknown 
parameters.  
\subitem The remaining vector and axialvector resonance contributions 
depend on eight unknown weak couplings. Various phenomenological
inputs can be used to fix them, but some remain unconstrained. 
In principle this part of the amplitude requires matching to the 
penguin short distance contribution. The requirement that matching
occurs successfully affords 
a way to further constrain the unknowns.  This is further discussed 
in Sect.~\ref{match}.
\subitem The terms involving scalar and pseudoscalar 
resonance exchange will also contain largely unconstrained 
couplings. At chiral order 
$e^2 p^2$, there are contributions from mass renormalizations on 
external legs and also from vertex-like corrections. The net 
effect at this order turns out to vanish. 
\item The low-energy parts of meson loop diagrams
(Fig.~\ref{fig:f3}(c)): 

We refer to these as the {\it unitarity} contributions. They 
constitute a genuine low-$Q^2$ effect distinct from that of the 
resonance component.  In Sect.~3.3 we shall describe such unitarity 
terms and provide a natural extension to all 
$Q^2$ scales, without introducing new parameters. 
\end{enumerate} 

Before proceding to a description of the calculation, 
we introduce a parameterization in terms of 
reduced amplitudes $\{ {\cal C}_{i}\}$ and 
$\{ {\overline W}_i \}$,
\beq
\delta {\cal A}_{i}^{\rm (em)}  =  \eta_{i} \frac{g_{8}M_{K}^{2}}
{F_{\pi}^{2}F_{K}} 
\frac{\alpha}{4 \pi} \, {\cal C}_{i} \quad {\rm and} \quad 
W_{i} =  \eta_{i} \frac{g_{8}M_{K}^{2}}
{F_{\pi}^{2}F_{K}} 
\frac{\alpha}{4 \pi} \, {\overline W}_i 
\eeq
with 
\beq
{\cal C}_{i}  \equiv \, \int_{0}^{\infty} \ d Q^2 \ 
{\overline W}_i (Q^2) 
\eeq
and $\eta_{+-} = \eta_{00} = \sqrt{2}$ and $\eta_{+0} = 1$.  
In addition, we partition each ${\cal C}$ amplitude into 
additive components as  
\beq
{\cal C}_{i}  = {\cal C}_{i}^{(e^2 p^0)} +   
{\cal C}_{i}^{\rm (mtchg)} + {\cal C}_{i}^{\rm (unty)} \ \ .
\label{comp}
\eeq
The matching component ${\cal C}_{i}^{\rm (mtchg)}$, 
encompassing the sum of the Born $+$ form factor and 
resonance contributions, is discussed in Sects.~3.1,3.2  
whereas the unitarity component 
${\cal C}_{i}^{\rm (unty)}$ is discussed in Sect.~3.3.
The contribution of each component to the full amplitude 
is given in Table 1 ({\it cf} Sect.~4).

\subsection{Born and Resonance Diagrams}\label{BVA}

The class of diagrams involving exchanges of the ground state 
pseudoscalar mesons and of the low-lying spin-one, spin-zero 
resonances generates contributions at 
order $e^2 p^0$ and at order $e^2 p^2$.  However, we already 
know the $e^2 p^0$ contributions because chiral 
symmetry relates the $K \rightarrow \pi \pi$ amplitudes 
to the  $K^{+} \rightarrow \pi^{+}$ matrix element and we have 
calculated this in Ref.~\cite{cdg}.  Therefore we focus 
on the $e^2 p^2$ piece in the following. 
We treat first in some detail the Born contributions and their 
corrections which arise from the insertion of meson form factors. 
Then we describe the parameter-dependent spin-one 
resonance terms and finally the spin-zero resonance terms.

\subsubsection{Born and Form Factor Contributions}
\begin{figure}[tbh]
\centering
\begin{picture}(200,150)  
\put(50,20){\makebox(100,120){\epsfig{figure=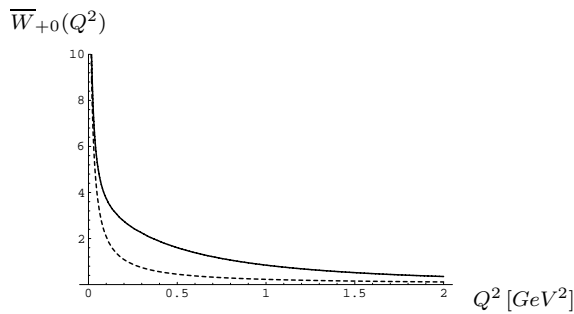,height=2in}}}
\put(180,30){\scriptsize{$Q^2 \, [GeV^2]$}}
\put(5,135){\scriptsize{${\overline W}_{+0} (Q^2)$}}
\end{picture}
\caption{Born (dashed) and Born plus form factor (solid) contributions. 
\label{fig:fig1}}
\end{figure}

The Born diagrams do not contribute to ${\overline W}_{00}(Q^2)$ (which 
involves only neutral particles) while giving nonzero contributions to 
both ${\overline W}_{+0} (Q^2)$ and ${\overline W}_{+-} (Q^2)$. 
For ${\overline W}_{+0} (Q^2)$ we find
\beq
{\overline W}_{+0} (Q^2) = \frac{3}{M_{K}^{2}} \, J (Q^2,M_{\pi}^{2})  \ \ ,
\eeq  
with 
\beq
J (Q^2,m^2) = \frac{Q^2}{6 m^2} \left[ \left( 1 + 
4 \frac{m^2}{Q^2} \right)^{\frac{3}{2}} - \left( 1 + 
6 \frac{m^2}{Q^2} \right) \right] .
\eeq  
This contribution is logarithmically divergent at high $Q^2$ and has 
an infrared $1/Q$ integrable singularity at $Q^2 = 0$. 
In addition, it is suppressed by a factor of 
$M_{\pi}^{2}/M_{K}^{2}$. This suppression is `accidental' in that 
it is not required by any symmetry at moderate or high values of
$Q^2$.  The result is shown as the dashed line 
in Fig.~\ref{fig:fig1}. 

\begin{figure}[tbh]
\centering
\begin{picture}(200,150)  
\put(50,20){\makebox(100,120){\epsfig{figure=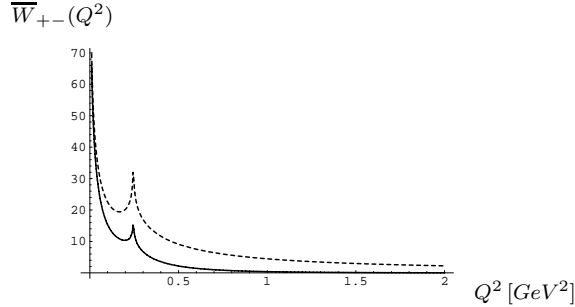,height=2in}}}
\put(180,30){\scriptsize{$Q^2 \, [GeV^2]$}}
\put(5,135){\scriptsize{${\overline W}_{+-} (Q^2)$}}
\end{picture}
\caption{Born (dashed) and Born plus form factor (solid) contributions.
\label{fig:fig2}}
\end{figure}

The Born contribution to ${\overline W}_{+-} (Q^2)$ is analytically more involved. 
Once one extracts the infrared divergent singularity~\cite{cdg4}, 
it reads
\beq
{\overline W}_{+-} (Q^2) = C (Q^2) + S_{1} (Q^2,M_{\pi}^{2}) - S_{2} (Q^2,M_{\pi}^{2}) \ . 
\eeq
The functions $C (Q^2)$, $S_{1} (Q^2,m^2)$ and $S_{2} (Q^2,m^2)$ are given 
in Appendix~\ref{ap1}, and we display ${\overline W}_{+-} (Q^2)$ as the dashed line 
in Fig.~\ref{fig:fig2}.   
Again, this contribution is logarithmically divergent at high $Q^2$.
The cusp is due to the singularity related to the coulombic rescattering. 

The set of Born diagrams, required by chiral symmetry,
 provides a good description of the very low $Q^2$ 
region, in which the photon `sees' only point-like pseudoscalars.
As $Q^2$ increases this is no longer true, and one needs to account 
for structure dependence in the couplings. In our model this is 
accomplished by introducing the low-lying resonances.

We consider first the diagrams involving pion and kaon electromagnetic 
form factors (saturated in this model by the vector meson resonances). 
This is a subclass of all diagrams required by chiral symmetry but has some 
nice features. It does not introduce any new parameters and improves the 
high-$Q^2$ behavior of the $\{ {\overline W}_{i} (Q^2)\} $ while having minimal effect 
on the model-independent Born contributions at low $Q^2$.
The results of this improved description are shown graphically 
in Figs. \ref{fig:fig1} and  \ref{fig:fig2} (solid lines). 
The anlytical expressions are
\beq
   {\overline W}_{+0} = \frac{M_{\rho}^{2}}{Q^2 + M_{\rho}^{2}}  
\bigg[ \frac{3}{M_{K}^{2}} \, J (Q^2,M_{\pi}^{2})  
 +  \frac{Q^2}{Q^2 + M_{\rho}^{2}} \,
 \tilde{J}(Q^2,M_{K}^{2}) \bigg]
\eeq  
and 
\beq
{\overline W}_{+-} = \left( \frac{M_{\rho}^{2}}{Q^2 + M_{\rho}^{2}} \right)^{2} \,
   C (Q^2) +  \frac{M_{\rho}^{2}}{Q^2 + M_{\rho}^{2}} \left(
 \frac{M_{\rho}^{2}}{Q^2 + M_{\rho}^{2}}  S_{1} (Q^2) -
  S_{2} (Q^2) \right)  \ ,
\eeq
where $\tilde{J} (Q^2,m^2)$ is defined in Appendix~\ref{ap1}.
Note that the new contribution to ${\overline W}_{+0}$ is not suppressed by 
$M_{\pi}^{2}/M_{K}^{2}$ and thus gives a substantial correction  
to the Born amplitude. In the case of ${\overline W}_{+-}$, however, the form 
factor contribution has simply the effect of softening the high-$Q^2$ 
behavior. 

In principle, given the convergence properties of the 
`Born + form factor' contributions, their integrations over 
$Q^2$ can be performed up to infinity. This contribution is 
dominated by the low and intermediate energy regions, where
the formalism is valid. This gives a first clean 
contribution to the $\{{\cal C}_{i}\}$ coefficients 
beyond the Born  approximation. 

\subsubsection{Resonance Contributions}
Our analysis contains two classes of resonances contributions, 
spin-one and spin-zero.  We consider each one separately 
in the following.  It turns out that the spin-zero 
contributions sum to zero, so that only the spin-one 
contributions are subject to the matching procedure of 
Sect.~3.2.

As noted earlier, chiral symmetry requires the presence of all
possible vector and axialvector resonance exchange diagrams. 
In principle these introduce to the $\{ {\overline W}_{i} (Q^2)\}$ 
a dependence on eight new parameters, describing the weak couplings  of 
resonances. Since the analytical expressions for this large class of 
contributions are rather lengthy and do not illuminate the underlying 
physics, we refrain from reporting them here. 
The only feature relevant for our discussion is the general
form, 
\beq 
\left. {\overline W}_{i} (Q^2) = \sum_{\alpha} \ \bar{g}_{\alpha} \ 
f_{\alpha}^{(i)} (Q^2) \ \ .
\right. 
\label{fcns}
\eeq 
Explicit calculation shows that the physical amplitudes actually depend 
 only on the seven parameters, 
\beq
\bar{g}_{1,2,5,6}^{(V)},\ \
\bar{g}_{5,6}^{(A)}, ~\  \bar{g}_{2}^{(A)}  - \bar{g}_{1}^{(A)} \ \ .
\eeq
Let us consider the high-$Q^2$ behavior of the 
functions $f_{\alpha}^{(i)} (Q^2)$ appearing in Eq.~(\ref{fcns}). 
Many of them go to a constant 
at high $Q^2$ or fall off as $1/Q^2$, thus leading to divergences 
in the integration process. 
Such behavior has already been observed in similar calculations 
of the electromagnetic mass shift of the kaon~\cite{dp,bu}.  
This simply means that the resonance dominance approximation can be 
trusted only up to some intermediate energy region and cannot be 
extended up to $Q^2 \rightarrow \infty$. 
In Sect.~\ref{match} we shall try to solve both these problems 
(proliferation of unknown parameters {\it and} high-$Q^2$ divergences) 
by requiring that the resonance amplitude contribution match 
the high-$Q^2$ behavior of the $\{ {\overline W}_{i}(Q^2)\}$.  

We consider next the spin-zero resonance contributions.  
 In the absence of electromagnetism, the tree level exchange of the 
scalar and pseudoscalar resonances
 contributes a major part of the $K \rightarrow \pi \pi$ amplitudes at
 chiral order $p^4$.~\cite{ekw} Dressing these diagrams with one
 virtual photon generates contributions to the amplitude 
$\delta {\cal A}_{i}^{\rm em}$
 at orders $e^2 p^2$ and $e^2 p^4$.  It is easy to convince
 oneself that diagrams with vertices coming from mass matrix
lagrangians, having already four powers of the pseudoscalar masses, 
  will contribute at order $e^2 p^4$ to $\delta
 {\cal A}_{i}^{\rm em}$. On the other hand, diagrams involving
 derivative vertices can give rise to effects of order $e^2 p^2$,
 which we are interested in. This happens through two classes of
 contributions:
\begin{enumerate}
\item mass renormalization on external legs, and 
\item vertex correction diagrams, with virtual photons inserted 
according to minimal coupling.
\end{enumerate}
Upon explicitly identifying and calculating these diagrams, we find
an exact cancellation between the two contributions.  This is identical
in nature to the one found in Ref.~\cite{cdg} for the Born
contributions at order $e^2 p^0$.  The explicit results (showing the
cancellation) can be found in Appendix~\ref{ap2}.

\subsection{The Matching Procedure}
\label{match}

As stated in the above discussions, the resonance exhange contribution 
provides a good description for the $\{ {\overline W}_{i} (Q^2)\}$ only up to 
some intermediate $Q^2$ region, beyond which the quark electroweak and 
strong interactions provide the correct framework.
Experience in similar hadronic calculations has shown that the transition 
or matching region occurs for $Q$ between 1.5 GeV and 
2 GeV (or $2 \le  Q^2~({\rm GeV}^2) \le 4$).  
The genuine short distance contributions were studied 
in the chiral limit in Ref.~\cite{cdg}. The outcome was that the 
short distance contribution to the $\{ {\overline W}_{i} (Q^2)\}$ is rather 
small compared to the long distance component. Corrections to the 
chiral limit cannot dramatically change this qualitative  picture. 
We can imagine assigning a $ 100 \% $
uncertainty to the short distance component around the central value given
 by the chiral limit calculation. Even in this case the long distance
 contribution would dominate and our ignorance of short distance
physics would not significantly alter the final 
answer. For our purposes, the most 
important feature emerging from this analysis is that for $Q^2 > \mu^2 $ the 
$\{{\overline W}_{i} (Q^2)\}$ can be set to zero, even if we do not know the
 details of this transition. 

On the other hand, for low and intermediate $Q^2$ we have reliable 
expressions for the $\{ {\overline W}_{i} (Q^2)\}$, {\it i.e.} the most general
 parametrization implied by chiral symmetry and the low-lying part of the 
hadronic spectrum. The only problem with these expressions is the presence of 
a large number of  resonance parameters unconstrained by phenomenology.
In what follows we shall present a set of reasonable theoretical 
constraints to be imposed on them.  
The underlying strategy is to use on the one hand the few  phenomenological
 inputs presently available, and on the other, to enforce the transition 
to the high-$Q^2$ region, meaning in our case that the 
$\{{\overline W}_{i} (Q^2)\}$ have to approach zero in the matching region.

\subsubsection{Physical Constraints on the
 $\bar{g}_{k}^{(V,A)}$}  

Although the $\{ \bar{g}_{k}^{(V,A)}\}$ of Eq.~(\ref{r6}) (see also
Eq.~(\ref{fcns})) are not predictable from a purely
theoretical approach, some information can be gleaned from the 
phenomenology of kaon decays and assorted theoretical requirements. 

The phenomenology of kaon decays, especially the radiative kaon decays,
allows in principle the extraction of a large number of order $p^4$
constants of the weak chiral lagrangian~\cite{ekw}. Assuming 
resonance dominance for these couplings (or whenever possible,
subtracting the short distance contribution) allows one to extract
information on the resonance coupling constants. The present
experimental situation does not, however, yet permit a complete
implementation of this program, as only limited information is
available. From $K \rightarrow 2 \pi, 3 \pi$ data and assuming
resonance saturation of the relevant ${\cal O} (p^4)$ counterterms,
one finds
\beq 
4 \bar{g}_{5}^{(V)} - \bar{g}_{6}^{(V)} = 0.43 \ \ ,    
\label{phen1}
\eeq 
with a 20$\%$ uncertainty associated with the extraction of 
${\cal O}(p^4)$ coupling constants.~\cite{ekw}  The 
$K^{+} \rightarrow \pi^{+} l^{+} l^{-}$ transition provides additional 
information. The decay amplitude depends on a parameter $w_{+}$, whose 
experimental value is $w_{+} = 0.89^{+ 0.24}_{-0.14}$.  
It receives both long and short distance contributions. Using resonance 
saturation at $\mu_{{\rm ChPT}} = M_{\rho}$ and including explicitly the
 penguin contribution we find
$$ \frac{1}{64 \pi^2} \left( 3 w_{+} - \log \frac{M_{\rho}^{2}}{M_{\pi} 
M_{K}} \right) - \frac{3 F_{\pi}^{2}}{2 M_{\rho}^2} - \frac{3}{72 \pi^2} 
\log \frac{m_{c}}{M_{\rho}} $$
\beq 
= \frac{\sqrt{2}}{M_{\rho}^{2}} \left[ \frac{F_{V}^{2}}{2} 
( \bar{g}_{6}^{(V)} - 2 \bar{g}_{5}^{(V)} ) - F_V G_V \bar{g}_{1}^{(V)} 
\right]   \ . 
\label{phen2}
\eeq
At present no other phenomenological constraints are available and we thus 
turn to the description of the theoretical ones. 

In the first place there are two conditions coming from the analysis 
performed in the chiral limit~\cite{cdg}. Let us recall the reasoning 
behind this.  In the chiral limit, using soft-pion methods, one can 
relate the $K \rightarrow \pi \pi$  amplitude to the off-shell 
$K^{+}$-to-$\pi^{+}$ matrix element. Moreover the invariant 
amplitude ${ \cal A}_{K^{+} \rightarrow \pi^{+}}$ is expressible as 
\beq
{ \cal A}_{K^{+} \rightarrow \pi^{+}} = \int_{0}^{\infty} \ d Q^2 
\ \bar{A}_{++} \, (Q^2) \ .
\eeq
We get one condition by demanding that  $\bar{A}_{++} \, (Q^2)$ 
vanish at infinity (no quadratic divergences). A second condition 
 comes from demanding that $\bar{A}_{++} \, (Q^2)$  have no short 
distance component, {\it i.e.}~that it vanish in the matching  
 region defined above. Together these amount to 
\beq 
\lim_{Q^2 \rightarrow \infty} \ \bar{A}_{++} \, (Q^2) = 0 \ \ \ \ \
\mbox{and} \ \ \ \ \ \bar{A}_{++} \, (\mu^2) = 0   \ \ .
\label{chc}
\eeq
Here we have introduced the matching scale $\mu$ and according to our 
previous discussion, let it vary between 1.5 GeV and 2 GeV. 
The following two constraints then emerge from Eq.~(\ref{chc}),
\beqa
& & {\bar g}_1^{\rm (V)} + {\bar g}_2^{\rm (V)} = 
f_V (\mu) \equiv {3\over 2 \sqrt{2}} \cdot 
{ F_\pi^2 \over F_V^2 - F_A^2 \cdot \displaystyle{{M_A^2 + \mu^2 \over 
M_V^2 + \mu^2 }}} \ \ ,
\nonumber \\
& & {\bar g}_1^{\rm (A)} - 3 {\bar g}_2^{\rm (A)} = 
f_A (\mu) \equiv {3\over 2 \sqrt{2}} \cdot
{ F_\pi^2 \over F_A^2 - F_V^2 \cdot \displaystyle{{M_V^2 + \mu^2 \over 
M_A^2 + \mu^2 }}} \ \ .
\label{c54}
\eeqa
Adopting the same argument, we require that ${\overline W}_{+-} (Q^2)$ 
and ${\overline W}_{00} (Q^2)$ vanish in the matching region,
\beq 
{\overline W}_{+-} \, (\mu^2) = 0 \ \ \ \ \
\mbox{and} \ \ \ \ \ {\overline W}_{+0} \, (\mu^2) = 0   \ \ .
\label{newc}
\eeq
We do not include the analogous condition for 
${\overline W}_{00} (Q^2)$ because 
this function, independent of any choice of the parameters,
is already very small in the matching region (it does not contain any 
term going to a constant for high $Q^2$). 
We believe that the conditions in 
Eqs.~(\ref{phen1}),(\ref{phen2}),(\ref{chc}),(\ref{newc}) form a 
consistent set of physical requirements and provide us
with a solid basis for any attempt to obtain a sensible answer for 
the so-called matching amplitudes $\{ {\cal C}_{i}^{\rm (mtchg)}\}$. 

\subsubsection{Results}

The above constraints are well motivated and reasonable, but
are not sufficient to completely fix all of the resonance parameters.
At this stage, we could use specific models of resonance behavior to 
estimate the remaining parameters and then accept the range of model 
dependence as an estimate of our uncertainty. In doing so, however, we 
have found that models generally give a rather small range of results. (The 
exception concerns ${\cal C}_{+0}^{\rm (mtchg)}$). The reason 
is that the matching constraint is more important than 
the remaining parameters. Therefore, rather than using particular
models we elect to follow a more model independent procedure of
allowing these remaining parameters to vary completely 
over their reasonable
physical range, and to use the resulting variation to estimate the
error bars for our result. 

\begin{center}
\begin{figure}
\vskip .1cm
\hskip 3.5cm
\epsfig{figure=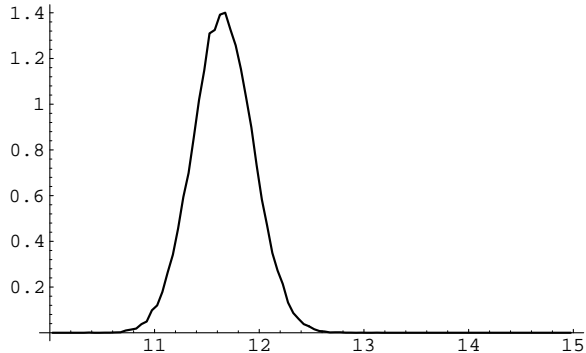,height=3in}
\caption{Probability density function for 
${\cal C}_{+-}^{\rm (mtchg)}$  .\hfill 
\label{fig:fig3}}
\end{figure}
\end{center}

The conditions described above imply a set of linear equations for the
parameters $\bar{g}_{i}^{(V,A)}$. In particular, we can express all 
the parameters in terms of just two coupling constants. That is, 
using  what we believe are well founded physical 
constraints, we select a two-dimensional hyperplane in the parameter 
space which we call the reduced parameter space. 
We chose as independent variables spanning this plane 
 $x = \bar{g}_{1}^{(V)}$ and $y = \bar{g}_{2}^{(A)} - g_{1}^{(A)}$. 
By looking at the structure of the constraints one discovers that 
the other parameters depend on $x,y$ and $\mu$ as 
\beq 
\bar{g}_{2}^{(V)} (x,\mu),  \ \
\bar{g}_{5,6}^{(V)} (x),  \ \ {\rm and} \ \   \bar{g}_{5,6}^{(A)} (x,y,\mu)  \ .
\eeq

\begin{center}
\begin{figure}
\vskip .1cm
\hskip 3.5cm
\epsfig{figure=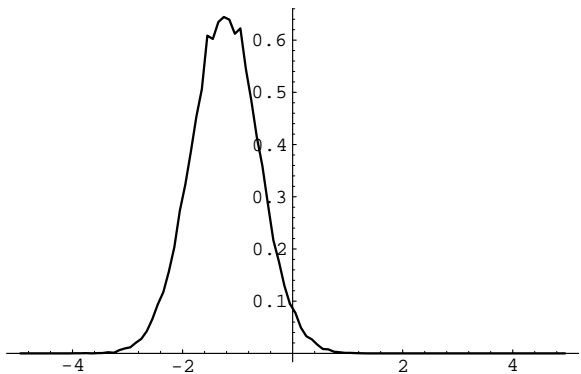,height=3in}
\caption{Probability density function for 
${\cal C}_{00}^{\rm (mtchg)}$  .\hfill 
\label{fig:fig4}}
\end{figure}
\end{center}

We are now in a position to determine the component 
${\cal C}_{i}^{\rm (mtchg)}$ of the full amplitude ${\cal C}_i $ 
which is determined via matching.  The construction described 
above allows us to express the predictions for each ${\cal
C}_{i}^{\rm (mtchg)}$ as a linear function of $x$ and $y$,  
\beqa 
& & {\cal C}_{+-}^{\rm (mtchg)} (x,y)  
=   (12.2 - 0.72 x + 0.02 y)  \pm \ |- 1.0 + 1
.35 x + 0.006 y| \ \ ,\nonumber  \\ 
& & {\cal C}_{+0}^{\rm (mtchg)} (x,y) 
=  (-9.2 + 11.47 x + 3.2 y)  \pm \ |- 4.3 +  
6.0 x + 1.0 y | \ \ , \nonumber  \\ 
& & {\cal C}_{00}^{\rm (mtchg)} (x,y)  
=  (- 0.0035 - 1.53 x )  \pm \  |- 0.0003 - 
0.42 x |  \ \ . \label{resxy} 
\eeqa
The uncertainties cited in Eq.~(\ref{resxy}) are associated with 
the matching procedure (the variation of the parameter $\mu$). 
Still, this leaves freedom to pick any value for $(x,y)$.  We can
further narrow our predictions by requiring that all the couplings
simultaneously (as a function of $(x,y)$) have a `natural' order of
magnitude, which can be shown to be ${\cal O}(1)$.  The existence of a
region in the $(x,y)$ plane such that this happens is not guaranteed a
priori and provides a good consistency check for our method. We call
this the {\em physical region} in the reduced parameter space.
Studying the explicit dependence of the parameters on $x$, $y$ and
$\mu$ we are lead to define the physical region as $x:
0.5 \rightarrow 1.5$ and $y: -1 \rightarrow 1$.  

\begin{center}
\begin{figure}
\vskip .1cm
\hskip 3.5cm
\epsfig{figure=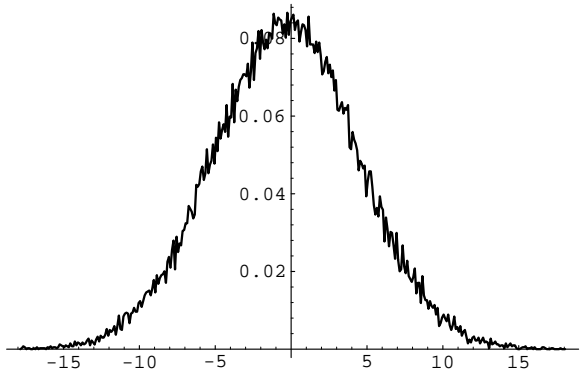,height=3in}
\caption{Probability density function for 
${\cal C}_{+0}^{\rm (mtchg)}$  .\hfill 
\label{fig:fig5}}
\end{figure}
\end{center}
 
A first qualitative conclusion can be
already drawn by looking at Eq.~(\ref{resxy}) with $(x,y)$ restricted
to the physical region and no further assumptions. The 
expressions show that ${\cal C}_{+-}^{\rm (mtchg)}$ depends 
very weakly on the choice of $(x,y)$ in this region, 
and thus we arrive at a good prediction for
this parameter.  ${\cal C}_{00}^{\rm (mtchg)}$ has a 
moderate dependence on $x$ 
and does not depend at all on $y$. This implies that the $K^{0}$
decay amplitudes can be predicted in our model with a reasonably small
uncertainty.  Problems arise in the expression for 
${\cal C}_{+0}^{\rm (mtchg)}$,
which displays a fairly strong dependence on $x$ and a moderate one 
on $y$. In this case, even confining ourselves to the physical
region we obtain a spread in the answers of about $100 \% $. The only
definite prediction emerging is that this contribution is not big. 

Quantitative estimates for our results and the attendant uncertainties
can be obtained by constructing probability distributions for the 
$\{{\cal C}_{i}^{\rm (mtchg)}\}$, by means of a 
survey of the parameter space. We scan the
region defined by $\{ -3 \leq x \leq 3, -3 \leq y \leq 3 \}$
using  gaussian distributions for the input parameters. The choice of the
parameters in the distributions is made in such a way to enhance 
contributions coming from the physical region. In view of this, we 
choose the central values as $x = 0.8$ and $y = 0$ and set variances 
equal to $0.4$.  The uncertainites cited in our results 
correspond to a $68\%$ probability.  Results for the 
$\{{\cal C}_{i}^{\rm (mtchg)}\}$ are given in the second row of Table 1.

\subsection{The Unitarity Diagrams}
\label{sub:uni}
Next we discuss in detail the class of diagrams
schematically represented in Fig.~3(c).  We
are interested in the nonlocal part of these diagrams, representing 
the genuine propagation of mesons at low energy.  The high 
momentum part of these diagrams produces (on general 
grounds~\cite{drv}) local effects that can be reabsorbed into the 
definition of the ${\cal O}(p^4)$ low energy constants.  
In our approach, however, the local
component at ${\cal O}(p^4)$ is implicitly contained in the
resonance exchange diagrams and would show up explicitly upon
expanding the resonance propagators. Keeping only the low momentum
part of the meson loops ensures that the different contributions we
are including in our calculation do not lead to double counting. Since 
the separation of local and nonlocal components in the meson loop
diagrams is not free of ambiguity, we shall be careful to 
describe and motivate our prescription in the following.\\

\subsubsection{Identification of the  ${\cal O}(e^2 p^2)$ Contribution}

\begin{figure}
\vskip .1cm
\hskip 1.2cm
\epsfig{figure=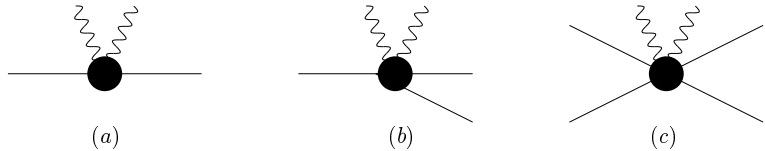,height=0.75in}
\caption{Two-photon insertions: (a) $T_{\mu\nu}$, (b) $V_{\mu\nu}$, 
(c) $S_{\mu\nu}$.\hfill 
\label{fig:fy}}
\end{figure}

Our task in the following is to identify the part in each 
meson loop diagram which, upon contracting the
photon legs, will lead to ${\cal O}(e^2 p^2)$ contributions.  
The loop contributions to $W_{\mu \nu}$ can be obtained 
by starting with any meson loop diagram which contributes to 
$K \rightarrow \pi \pi$ and attaching two photons in all 
possible ways. We focus first on the subclass of diagrams obtained by 
attaching the following two-photon insertions,
\beqa
T_{\mu \nu} (p,q), \ \ \ \ & V_{\mu \nu} (p_{i},q), & \ \ \ \ S_{\mu
\nu} (p_{i},q), 
\eeqa
as represented in Fig.~\ref{fig:fy}.

\begin{figure}
\vskip .1cm
\hskip 3.0cm
\epsfig{figure=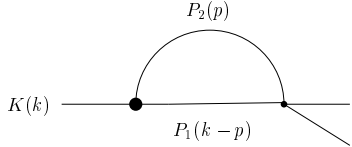,height=0.75in}
\caption{Loop diagram with internal particles $P_1$ and $P_2$.\hfill
\label{fig:fz}}
\end{figure}

For definiteness let us refer to the bare topology of Fig.~\ref{fig:fz}.
In this case one can insert $T_{\mu \nu} (p,q)$ on internal and external
 legs, $V_{\mu \nu} (p_{i},q)$ in the weak vertex and $S_{\mu
\nu} (p_{i},q)$ in the strong vertex. The external leg insertions will 
generate wavefunction and mass renormalizations. The other insertions will  
give rise to diagrams like 
\beqa
& & D_{\rm mass} = \int d^{4} p ~ \frac{V_{w} (p_{i}) \ V_{s} (p_{i})}{ 
\left[(k - p)^2 - M_{P_{1}}^2 \right] \left[p^2 - M_{P_{2}}^2 \right]^2} 
\int d^{4}q~ D^{\mu\nu}(q)~ T_{\mu \nu} (p,q)  \ ,\nonumber \\
& & D_{\rm weak} = \int  d^{4} p \ \frac{ V_{s} (p_{i})}{ 
\left[(k - p)^2 - M_{P_{1}}^2 \right] \left[p^2 - M_{P_{2}}^2 \right]} 
\int d^{4}q~ D^{\mu\nu}(q)~  V_{\mu \nu} (p_{i},q)  \ ,
\nonumber \\
& & D_{\rm strong} = \int \, d^{4} p \ \frac{V_{w} (p_{i})}{ 
\left[(k - p)^2 - M_{P_{1}}^2 \right] \left[p^2 - M_{P_{2}}^2 \right]} 
\int d^{4}q~ D^{\mu\nu}(q)~  S_{\mu \nu} (p,q)  \ . \nonumber \\
\label{u1} 
\eeqa 
We wish to isolate the dominant contributions at low momentum 
(small $p$).  Therefore we Taylor expand each tensor insertion 
around $p_{i} = 0$ (in addition we must expand each 
coefficient of the 
Taylor series in powers of the pseudoscalar meson masses; for notational 
convenience we don't explicitly display this step). 
Considering for example the self-energy insertion, one has 
\beq
T_{\mu \nu} (p,q) = T_{\mu \nu} (0,q) + 
p_{\alpha} {\partial T_{\mu \nu} \over \partial p_\alpha}(0,q)  
+ \frac{1}{2!} ~ p_{\alpha} p_{\beta}
{\partial^2 T_{\mu \nu} \over \partial p_{\alpha} \partial p_{\beta}} 
(0,q) + \dots \ .
\eeq 
Evaluation of the integrals in Eq.~(\ref{u1}) can be done term-by-term  
in the series.  The analysis of each term is very simple.  The 
tensor structure $\partial^{n} / \partial p^{n} T_{\mu \nu} (p,q)
|_{p=0}$ factorizes out of the integration over $p$, and the two-photon
insertion is replaced by a meson vertex of order $p^{n}$. 
This makes power counting transparent --- after
contracting the photon legs it is easy to realize that {\em only the
 first term} in the above expansion produces an effect of order 
$e^2 p^2$ in the kaon ampliude. In the example considered one has 
\beq
D_{\rm mass} =  \int \, d^{4} p \ \frac{V_{w} (p_{i}) 
\ V_{s} (p_{i})}{ \left[(k - p)^2 -
M_{P_{1}}^2 \right] \left[p^2 - M_{P_{2}}^2 \right]^2} 
\int \, d^{4} q \
D^{\mu \nu} (q) T_{\mu \nu} (0,q) \ \ + \dots 
\label{massins}
\eeq

This procedure allows us to identify and interpret the relevant 
contributions at order $e^2 p^2$. 
The integral of $T_{\mu \nu}(0,q)$, weighted by the photon propagator 
in the above expression, is exactly the expression for the electromagnetic
self-energy of a charged meson in the chiral limit. Thus the insertion 
of $T_{\mu\nu}$ in a loop diagram reproduces the effect of inserting
the electromagnetic mass difference into such a loop. 
Eq.~(\ref{massins}) then  represents the meson diagram 
of Fig.~\ref{fig:fz} with a mass shift insertion on the $P_{2}$ 
intermediate leg.
Analogously, insertions of $V_{\mu \nu}$ and $S_{\mu \nu}$ yield 
Fig.~\ref{fig:fz} but with the weak and strong
vertices replaced (respectively) by constant vertices of order $e^2 p^0$, 
proportional to the chiral couplings  $g_{emw}$ and $g_{ems}$~\cite{cdg2}.
In other words these contributions are the counterparts 
to what were called {\em implicit} diagrams in the ChPT 
calculation of Ref.~\cite{cdg2}. Their presence in the dispersive 
matching model is welcome because they provide imaginary 
parts to the amplitudes, ensuring at this order the behavior 
 required by unitarity. These expressions when parametrized in 
terms of $g_{\rm emw}$ and $\delta M_{\pi}^{2}$, are identical to
those obtained in ChPT (see Eq.~(30) of Ref.~\cite{cdg2}). 
Corresponding to these contributions will, of course, also be 
nonvanishing real parts, whose treatment is the subject of the next 
subsection. 

What becomes of the class of diagrams having separated photon 
vertices?  The basic result is that they start
contributing to $K \rightarrow \pi \pi$ amplitudes at order $e^2
p^4$. A short argument for this is as follows.  Upon contracting the
photon legs, it is easy to recognize that these loop diagrams have the
following peculiarity --- their intermediate states always involve a
photon (they contain one photon plus one or two pseudoscalar
mesons). Let us now consider the diagrams as analytic functions of the
external four momenta and analyze their imaginary parts as obtained by 
using the cutting equations.  The above observation on the structure
of the intermediate states, together with the form of the lowest order
vertices and the phase space, implies that the imaginary part of these
diagrams is of order $e^2 p^4$. The long distance portion of the real
part of the loop, which is all that we are interested in here, will then
appear at the same chiral order.

\subsubsection{Real-parts of the Unitary Amplitudes}
In the previous subsection we showed that the relevant part of the unitarity
contributions at order $e^2 p^2$ can be calculated with a simple recipe: 
the photon insertion factorizes out and one is left with the calculation  
of meson loop integrals with mass insertions on internal or external 
legs and weak or strong vertices replaced by constant vertices. 
As we have stated in the introduction to this section, we want
 to keep only the low energy part of these meson loops, the one that 
cannot be mimicked by any local counterterm or resonance-exchange diagram. 
We can best describe this procedure using a simple case, the two-pion loop, 
which also turns out to be the most relevant for the physics.
 The extension to all other diagrams is then straightforward. 

The basic function entering the description of two-pion loops is 
$J_{\pi \pi} (s)$, which is given in dimensional regularization by 
\beq 
J_{\pi \pi} (M_{K}^{2}) = \frac{1}{(4 \pi)^2} \, 
\left[ D_{\epsilon} +  \log \frac{\nu^2}{M_{\pi}^{2}} + 1 + 
\beta \log \left( \frac{\beta - 1}{\beta + 1} \right) \right] \ \ ,
\label{u2}
\eeq 
where 
\beq 
D_{\epsilon} \equiv \left(\frac{2}{4 - d} - 
\gamma + \log 4 \pi + 1 \right) \ \ , 
\eeq
$\nu$ is the scale parameter introduced in dimensional regularization and 
$\beta$ is the pion velocity in the kaon rest frame. 
The divergent piece and the scale dependent logarithm 
in Eq.~(\ref{u2}) are clearly 
local effects. On the other hand, the last term  and the 
$\log M_{\pi}^{2}$ term are associated with the low energy meson
 propagation. Finally, an explicit cutoff calculation shows that  
the additive factor of one has to be included in the long-distance part. 
These considerations lead us to introduce a separation scale 
$\Lambda_{s}$ such the short and long distance parts are defined as
\beq 
J_{\pi \pi}^{\rm (SD)}  (M_{K}^{2}) \equiv \frac{1}{(4 \pi)^2} \, 
\left( D_{\epsilon} +  \log \frac{\nu^2}{\Lambda_{s}^{2}}  \right) \ \ ,
\label{u3}
\eeq 
\beq 
J_{\pi \pi}^{\rm (LD)} (M_{K}^{2}) \equiv \frac{1}{(4 \pi)^2} \, 
\left(  \log \frac{\Lambda_{s}^{2}}{M_{\pi}^{2}} + 1 + 
\beta \log \left( \frac{\beta - 1}{\beta + 1} \right) \right) \ \ .
\label{u4}
\eeq 
There is an inherent ambiguity in the separation scale 
$\Lambda_{s}$ which cannot realistically be assigned a unique value.  
Therefore we let it range between $M_{K}$ and $M_{\rho}$, 
associating the corresponding variation in the result as the theoretical 
uncertainty. These unitarity corrections come with moderate 
uncertainties except for the case of ${\cal C}_{00}$ 
({\it cf} Table 1).

\subsection{EM Corrections to the Isospin Amplitudes}
Let us consider some phenomenological consequences of our 
analysis.  We refer the reader to Sect.~2 and to Sect.~4.3 
of Ref.~\cite{cdg2} for an introduction to formalism used 
in the following.  In the presence of electromagnetism, the 
amplitudes involving 
charged particles (${\cal A}_{+-}$ and ${\cal A}_{+0}$) contain infrared
singularities.  For each such amplitude, the infrared singularity can, 
on general grounds, be isolated in an exponential factor that 
multiplies an infrared-finite amplitude which can itself be expressed 
as an expansion in powers of alpha.  Upon considering 
the emission of soft photons with energy up to some experimental 
scale $\omega$, the infrared divergences disappear from the decay 
rate expressions, leaving $\omega$-dependent 
factors $G_{+-} (\omega)$, $G_{+0} (\omega)$ which 
multiply the square moduli of the infrared-finite amplitudes. This
process has been explicitly described in Sect.~4.3 in Ref.~\cite{cdg2}
for the $K^{0} \rightarrow \pi^{+} \pi^{-}$ mode. 

Starting from the
infrared-finite amplitudes in the {\em charge} basis, we can define
the would-be isospin amplitudes from the following linear
combinations, 
\beqa
{\cal A}_{0} &  = & \frac{2}{3} {\cal A}_{+-} + \frac{1}{3} {\cal
A}_{00} \ \ , 
\nonumber \\
{\cal A}_{2} &  = & \frac{\sqrt{2}}{3} ( {\cal A}_{+-} - 
{\cal A}_{00} ) \ \ , 
\label{isospin}  \\ 
{\cal A}_{2}^{+} & = & \frac{2}{3} {\cal A}_{+0} \ \ . \nonumber 
\eeqa 
In the absence of electromagnetism and any other isospin breaking
interaction, we then have ${\cal A}_{2} = {\cal A}_{2}^{+}$, 
and the amplitudes
of Eq.~(\ref{isospin}) truly represent transitions to pure isospin 
states. Using the same logic one can perform an analysis of the 
unitarity condition~\cite{cdg4}, leading to the following parametrization 
of the  $K \rightarrow \pi \pi$ infrared finite amplitudes, 
\beqa
{\cal A}_{+-} &=& \left( A_0 + \delta A_0^{\rm em} \right)
e^{i(\delta_0 + \gamma_0)} + { 1 \over \sqrt{2}} \left( A_2 +
\delta A_2^{\rm em}  \right)
e^{i(\delta_2 + \gamma_2)}
\ , \nonumber \\
{\cal A}_{00} &=& \left( A_0 + \delta A_0^{\rm em}  \right)
e^{i(\delta_0 + \gamma_0)} - \sqrt{2} \left( A_2 +
\delta A_2^{\rm em}  \right)
e^{i(\delta_2 + \gamma_2)} \ \ ,
\label{a6} \\
{\cal A}_{+0} &=& {3 \over 2} \left( A_2
+ \delta A_2^{+{\rm em}}  \right)
e^{i (\delta_2 + \gamma_2')} \ \ .
\nonumber
\eeqa
The calculation performed in this paper gives us knowledge of the 
$\{ \delta A_I^{\rm em}\}$.  We find 
the shifts in the isospin amplidudes to be 
\beqa
& & \delta A_0^{\rm em} =  {\sqrt{2} g_{8} M_{K}^{2} \over F_{K} F_{\pi}^{2}} 
{\alpha \over 4 \pi} \, 
\left( \frac{2}{3} {\cal C}_{+-} + \frac{1}{3}  {\cal C}_{00} \right)  =
(0.0253  \pm 0.0072 ) \, 10^{-7} M_{K^{0}}  \ , \nonumber \\
& & \delta A_2^{\rm em} = \frac{\sqrt{2} g_{8} M_{K}^{2} }{F_{K} F_{\pi}^{2}} 
\frac{\alpha}{4 \pi} \, 
\frac{\sqrt{2}}{3} \left( {\cal C}_{+-} -  {\cal C}_{00} \right)  =
( 0.0118 \pm 0.0063 ) 10^{-7} \,  M_{K^{0}}   \ ,\nonumber \\
& & \delta A_2^{+ {\rm em}} =   \frac{g_{8} M_{K}^{2} }{F_{K} F_{\pi}^{2}} 
\frac{\alpha}{4 \pi} \, \frac{2}{3} {\cal C}_{+0}  = 
 - (0.0080  \pm 0.0088 ) 10^{-7} \,  M_{K^{0}}  \ \ . \label{isospin1} 
\eeqa
In our numerical evaluation we have used a value for $g_{8}$ 
obtained from a fit to data not including radiative
corrections.~\cite{kmw}  This introduces an ambiguity in 
$g_{8}$ of order $\alpha$ which affects $\delta A_I^{\rm em}$ at 
order $\alpha^2$ and thus is beyond the accuracy we are working at.  
As a byproduct we obtain also the effective $\Delta I = 5/2$ amplitude,
\beqa
{\cal A}_{5/2} &=& \frac{\sqrt{2} g_{8}}{F_{K} F_{\pi}^{2}} 
\frac{\alpha}{4 \pi} \, M_{K}^{2} \, {\sqrt{2} \over 5} \, 
\left(  {\cal C}_{+-} -  {\cal C}_{00} -  {\cal C}_{+0} \right) 
\nonumber \\
&=& ( 0.0137 \pm 0.0097 ) \cdot 10^{-7} M_{K^{0}} \ \ .
\eeqa

\section{Conclusions}
The problem of determining electromagnetic corrections to 
nonleptonic kaon decay is a formidable one and has long 
resisted understanding.  In this paper we have employed 
a `dispersive matching' approach which provides a framework 
that is, in principle, general and model-independent. This 
dispersive setting was first advocated by Cottingham~\cite{cott} 
and has been recently employed in Ref.~\cite{dp}.  At a 
practical level, however, a rigorous implementation of this program 
is plagued by a lack of sufficient input data.  
We have been able to overcome this obstacle 
by pointing out (on rather general grounds) how long range 
and intermediate range processes are expected to dominate the 
physics and then performing a calculation which 
incorporates the relevant ground state and resonance degrees 
of freedom.  All possible tree-level amplitudes and a subset of loop 
amplitudes are taken into account.  The latter component ensures 
that our amplitudes have the imaginary parts required by 
unitarity.  

\begin{center}
\begin{tabular}{c|ccc}
\multicolumn{4}{c}{Table~1: {The $\{{\cal C}_{i}\}$ Amplitudes }}
\\ \hline\hline
&  ${\cal C}_{+-}$ & ${\cal C}_{00}$  &  ${\cal C}_{+0}$ \\
\hline 
$e^2 p^0$ & $- 3.3 \pm 1.7$   &    $ 0 $     & $ -3.3 \pm 1.7$ \\
$e^2 p^2$ (Matching)  & $11.6 \pm 0.3$   &  $-1.3 \pm 0.7$ & $  - 0.4 \pm 4.5 
    $    \\
$e^2 p^2$ (Unitarity) & $6.5 \pm 1.5 $    &  $3.1 \pm 1.4$  & $- 3.4 \pm 1.2$ \\
  Total        & $ 14.8 \pm 3.5  $  &  $1.8 \pm 2.1$ & $- 7.1 \pm 7.4$   \\
\hline 
ChPT  &  $14.1 \pm 12.5 $ & $0.9 \pm 6.7$  &  $-4.2 \pm 4.6$ \\
\hline 
\end{tabular}
\end{center}

Contributions to the $\{ {\cal C}_{i}\}$ are shown in Table 1.   
The first row displays terms of order $e^2 p^0$ calculated in 
Ref.~\cite{cdg}. The equal values for ${\cal C}_{+-}$ and ${\cal C}_{+0}$ 
are due to the absence of a $\Delta I = 5/2$ component at lowest order. 
The second and third rows display terms at order $e^2 p^2$, 
arising from the analyses done in Sects.~\ref{BVA}-3.2 and in 
Sect.~\ref{sub:uni}. 
The fourth row summarizes the total result obtained 
within the dispersive matching approach.  
For comparisons sake, we also cite the ChPT results in the final row.  
We obtain EM corrections whose 
central values are in reasonable accord with our earlier ChPT 
calculation~\cite{cdg2} but whose theoretical 
uncertainties are substantially smaller for the 
$K^0 \to \pi^+ \pi^-$ and $K^0 \to \pi^0 \pi^0$ modes.  
Only the $K^{+} \rightarrow \pi^{+} \pi^{0}$ determination 
produces a less precise value.  It is not hard to 
recognize the reason for this.  In our
language, ${\cal C}_{+-}$ is dominated by the low-$Q^2$ Born
contribution. The resonance contribution, important at intermediate
$Q^2$, introduces only a moderate uncertainty. On the other hand, 
for ${\cal C}_{+0}$ the low-$Q^2$ contribution is very small, 
being suppressed by a factor $M_{\pi}^{2}/M_{K}^{2}$ (see Sect.~\ref{BVA}). 
${\cal C}_{+0}$ is thus dominated by intermediate $Q^2$ effects, 
which are plagued by a substantial uncertainty that our constraints 
have not completely eliminated.  We could turn to model-dependent
frameworks to attempt to narrow the quoted error bars. However, 
this apparent improvement would likely be illusory, since our 
understanding of models is too weak for any specific model to be 
trusted in a calcuation such as this. Thus we feel that our quoted 
error bars are a reasonable measure of present uncertainties. Note 
however, that the uncertainty in  ${\cal C}_{+0}$ is not
much of a problem because of the overall smallness of the effect.

A key result of our calculations is that the electromagnetic corrections to the
weak amplitudes are smaller than naive estimates might indicate. Part of the
reason is the partial cancellation in the leading chiral transition that we 
detailed in Ref.~\cite{cdg}. In addition, only about a third of the overall
electromagnetic effect goes into a modification of the I=2 final state - the
rest is harmless as it contributes to 
the much larger I=0 final state amplitude.
Although the work done here constitutes a crucial step 
in our study of EM corrections to nonleptonic kaon decay, 
there remain several additional issues which we shall address 
in a future publication.~\cite{cdg4}  Chief among these is 
how to correctly extract the electromagnetically 
corrected $K \to \pi\pi$ amplitudes from experimental data.  
We shall discuss the underlying theory in some detail, 
as well as suggesting the proper procedure to be followed 
in the experimental analysis.  Another topic to be covered, 
of great current interest in 
studies of CP violation, involves the ratio $\epsilon'/\epsilon$.
The calculation done here leads to a value for the EM correction 
to $\epsilon'/\epsilon$ (commonly denoted as 
$\Omega_{\rm EM}$).  We shall also provide an improved  
determination of the phase of $\epsilon'/\epsilon$.

The research described here was supported in part by the 
National Science Foundation.  One of us (V.C.) acknowledges 
support from M.U.R.S.T.

\eject

\appendix
\section{Loop functions}
\label{ap1}

In this appendix we give the analytic form of the functions entering 
in the Born term of $W_{+-} (Q^2)$. It is convenient to express them in 
terms of four simpler functions arising from the integration over the 
angular variables,
\beqa
C(Q^2) & = &  Q^2 \left[ F_{3} (Q^2) + 3 F_{4} (Q^2) + 
\frac{Q^2 F_{4} (Q^2) + F_{1} (Q^2,M_{K}^{2}) 
}{2 M_{K}^{2}} \right] \ , \nonumber \\
S_{1} (Q^2,m^2) & = & 2 \left( F_{1} (Q^2,m^{2}) - \beta^{2}  
F_{2} (Q^2,m^2) \right) \ , \nonumber \\ 
S_{2} (Q^2,m^2) & = & 2 \beta^2 \left( F_{1} (Q^2,m^2) - 
F_{2} (Q^2,m^2) \right) \ , \nonumber \\
\tilde{J} (Q^2,m^2)  & = & 2 F_{1} (Q^2,m^{2}) + F_{2} (Q^2,m^{2}) 
\ \ .
\label{app1}
\eeqa
The $F_{i}$ are 
\beqa
F_{1} (Q^2,m^2) & = &  \frac{1}{2 m^2} \,  \left( -1 + \sqrt{1 +
 4 \frac{m^2}{Q^2}} \right) \ , \nonumber \\
F_{2} (Q^2,m^{2}) & = &   \frac{Q^2}{8 m^4} \left[ \left( 1 +
 \frac{2 m^2}{Q^2} \right) - \sqrt{1 - 4 \frac{m^2}{Q^2}} \right] 
\ , \nonumber \\
F_{3} (Q^2) & = & \frac{1}{\beta M_{K}^{2} Q^2} \left[
\log \left( \frac{1 - \beta}{1 + \beta} \right) 
- \log  \bigg|  \frac{1 - \beta \sqrt{1 + \frac{4 M_{\pi}^2}{Q^2}}}{1 + 
\beta   \sqrt{1 + \frac{4 M_{\pi}^2}{Q^2}} } \bigg| \right] 
\ , \nonumber \\
F_{4} (Q^2) & = & \frac{1}{2 M_{\pi}^{2} Q^2} \left(  1 - 
\sqrt{\frac{Q^2}{Q^2 + 4 M_{\pi}^{2}}} \right) \ \ .
\label{app2}
\eeqa

\section{Scalar and pseudoscalar resonance contribution}\label{ap2}
Let us define the parameter $I_{\rm EM}$ as follows, 
\beq
I_{\rm EM} = \frac{3 \alpha}{4 \pi} \, \int_{0}^{\Lambda^2} \, 
d Q^2 \ \ .
\eeq
The scalar and pseudoscalar resonance contribution to the vertex-correction 
diagrams at order $e^2 p^2$ is given by the following expressions 
for ${\cal A}_{+-}$,  
\beqa
 & & \frac{{\cal A}_{+-}}{I_{\rm EM}} 
=  \frac{4 \sqrt{2}}{3 F_{\pi}^2 F_{K}} 
\, c_{m} \, ( 2 g_{4}^{S} + g_{6}^{S} ) \, 
\frac{M_{K}^{2} - M_{\pi}^{2}}{  M_{S}^{2}} 
-  \frac{4 \sqrt{2}}{F_{\pi}^2 F_{K}} 
\, \tilde{g}_{4}^{P}  \tilde{d}_{m} \, 
\frac{ M_{K}^{2}}{ M_{P}^{2}}
\nonumber \\
& &  - \frac{4 \sqrt{2}}{3 F_{\pi}^2 F_{K}} \, c_{d} \, \left[ 
( 4 M_{K}^{2} - M_{\pi}^{2} ) g_{1}^{S} + 
3 ( M_{K}^{2} + \frac{1}{2} M_{\pi}^{2} ) g_{2}^{S} \right] 
\frac{1}{M_{S}^{2}} 
\nonumber \\
& &  +  \frac{\sqrt{2}}{F_{\pi}^2 F_{K}} 
\, \tilde{g}_{2}^{S}  \tilde{c}_{m} \, 
\frac{4 M_{K}^{2} + 2 M_{\pi}^{2}}{ M_{S_{1}}^{2}} 
+  \frac{4 \sqrt{2}}{F_{\pi}^2 F_{K}} 
\, \tilde{g}_{1}^{S}  \tilde{c}_{d} \, 
\frac{ M_{K}^{2} - M_{\pi}^{2}}{ M_{S_{1}}^{2}}  \ \ , 
\label{b2} 
\eeqa
and for ${\cal A}_{+0}$,
\beqa
& &  \frac{{\cal A}_{+0}}{I_{\rm EM}} =  - \frac{4}{3 F_{\pi}^2 F_{K}} 
\, c_{m} \, g_{4}^{S} \, 
\frac{M_{K}^{2} - M_{\pi}^{2}}{ M_{S}^{2}} 
\nonumber \\
& & - \frac{4}{ F_{\pi}^2 F_{K}} \, c_{d} \, \left[ 
  M_{K}^{2} g_{1}^{S} + ( M_{K}^{2} + \frac{1}{2} M_{\pi}^{2} ) g_{2}^{S} + 
  M_{\pi}^{2}  g_{3}^{S}  \right] 
\frac{1}{ M_{S}^{2}} 
\nonumber \\
& &  +  \frac{1}{F_{\pi}^2 F_{K}} 
\, \tilde{g}_{2}^{S}  \tilde{c}_{m} \, 
\frac{4 M_{K}^{2} + 2 M_{\pi}^{2}}{ M_{S_{1}}^{2}} 
-  \frac{4}{F_{\pi}^2 F_{K}} 
\, \tilde{g}_{4}^{P}  \tilde{d}_{m} \, 
\frac{ M_{\pi}^{2}}{  M_{P}^{2}} \  .
\eeqa
The mass renormalization effect is given by the negative of 
this expression, 
with $I_{\rm EM}$ replaced by $\delta M_{\pi}^{2}$. 
However, the long distance contribution of $\delta M_{\pi}^{2}$ is just 
given by $I_{\rm EM}$ and thus these terms cancel each other. 
We neglect any residual intermediate energy component that 
may occur.  


\begin{thebibliography}{99}
\def\NPB #1 #2 #3 {Nucl.~Phys.~{\bf#1},\ (#2)\ #3}
\def\PLB #1 #2 #3 {Phys.~Lett.~{\bf#1},\ (#2) #3}
\def\PR #1 #2 #3 {Phys.~Rep.~{\bf#1},\ (#2) #3}
\def\PRD #1 #2 #3 {Phys.~Rev.~{\bf#1},\ (#2) #3}
\def\PRL #1 #2 #3 {Phys.~Rev.~Lett.~{\bf#1},\ (#2) #3}
\def\RMP #1 #2 #3 {Rev.~Mod.~Phys.~{\bf#1},\ (#2) #3}
\def\ZP #1 #2 #3 {Z.~Phys.~{\bf#1},\ (#2) #3}
%

\bibitem{cdg2} {\it Electromagnetic Corrections
to $K \to \pi\pi$ I -- Chiral Perturbation Theory},
V. Cirigliano, J.F. Donoghue and E. Golowich, hep-ph/9907341.

\bibitem{cdg} V. Cirigliano, J.F. Donoghue 
and E. Golowich, Phys. Lett. {\bf B450}  (1999) 241.

\bibitem{dp} J.F. Donoghue and A. Perez, Phys. Rev.
{\bf D55} (1997) 7075.

\bibitem{weakres} G. Ecker, J. Gasser, A. Pich and E. de Rafael, Nucl.
Phys. {\bf {B321}} (1989) 311; 

\bibitem{ekw} G. Ecker, J. Kambor, and D. Wyler,
Nucl. Phys. {\bf B394} (1993) 101.

\bibitem{cdg4} {\it $K \to \pi\pi$ Final State Phases in
the Presence of Electromagnetism}, V. Cirigliano, J.F. Donoghue
and E. Golowich, in preparation.

\bibitem {bu} R. Baur and R. Urech, Nucl. Phys.
{\bf {B499}} (1997) 319.

\bibitem {drv} J.F. Donoghue, C. Ramirez, G. Valencia,  Phys. Rev.
{\bf {D 39}} (1989) 1947.


\bibitem {kmw} J. Kambor, J. Missimer and D. Wyler,
Phys. Lett. {\bf B261} (1991) 496.

\bibitem{cott} W.N. Cottingham, Ann. Phys. {\bf 25} (1963) 424.

\end{thebibliography}
\end{document}